\RequirePackage{fix-cm}
\documentclass[twocolumn]{svjour3}          %
\smartqed  %
\usepackage[square,sort,comma,numbers]{natbib}
\usepackage{listings}
\usepackage{graphicx}
\usepackage{amsmath}
\usepackage{subfigure} 
\usepackage{booktabs}
\usepackage{multirow}
\usepackage{listings}
\usepackage{algorithm}  
\usepackage{algpseudocode}  
\usepackage{amsmath}  
\usepackage{bm}
\makeatletter
\renewcommand{\maketag@@@}[1]{\hbox{\m@th\normalsize\normalfont#1}}%
\makeatother
\newcommand{\tabincell}[2]{\begin{tabular}{@{}#1@{}}#2\end{tabular}}

\journalname{Structural and Multidisciplinary Optimization}
\renewcommand{\makeheadbox}{}

\begin{document}

\title{Parallel Multi-Fidelity Expected Improvement Method for Efficient Global Optimization%
}

\author{Zhendong Guo$^*$\and
Qineng Wang \and
Liming Song\and
Jun Li}
\institute{Zhendong Guo$^*$(corresponding author)  \at
              Institute of Turbomachinery, Xi'an Jiaotong Universiy, Xi'an 710049, China \\
	       \email{ericzhendong@163.com}     
	 \and
	 Qineng Wang \at
              	Institute of Turbomachinery,
	Xi'an Jiaotong University,
	Xi'an 710049, China \\
              \email{zhet1997@stu.xjtu.edu.cn}           
              \and
           Liming Song \at
              Institute of Turbomachinery,
	Xi'an Jiaotong University,
	Xi'an 710049, China \\
              \email{songlm@mail.xjtu.edu.cn}     
                \and 
          Jun Li \at
              Institute of Turbomachinery,
	Xi'an Jiaotong University,
	Xi'an 710049, China \\
              \email{junli@mail.xjtu.edu.cn}  
}
\date{}
\maketitle
\begin{abstract}
Multi-Fidelity optimization (MFO) has received extensive attentions in engineering design,   
which resorts to augmenting the small number of expensive high-fidelity (HF) samples by a large number of low-fidelity (LF) but cheap samples to improve the optimization performance. 
A key factor that influences the effectiveness of MFO is how to adaptively assign samples for HF and LF simulations in the iteration process.
To address such sample assignment issue in MFO, we propose a new infill criterion named as Filter-GEI, which imposes an adaptive filter function on top of the generalized expected improvement (GEI) acquisition function.
In particular, by taking the correlations between HF and LF models into account, the Filter-GEI can efficiently allocate HF and LF samples to achieve a good balance in between the local and global search. 
Furthermore, considering parallel computing, the Filter-GEI infills multiple HF and LF samples in each iteration, which can further improve its efficiency as computing power increases. 
Through tests on five mathematical toy problems and one engineering problem for the turbine blade design, the effectiveness of the proposed algorithm has been well demonstrated. 
\keywords{Multi-Fidelity optimization \and Co-kriging surrogate \and Generalized expected improvement \and Efficient global optimization \and Infill-sampling criterion}
\end{abstract}
\section{Introduction}
\par 
The efficient global optimization (EGO) is well known to be sample efficient~\cite{jones_efficient_1998}, which has achieved tremendous success in solving engineering problems~\cite{song_research_2016}. 
 Basically, an EGO algorithm consists of four steps as 
(1) initial sampling; 
(2) surrogate modeling; 
(3) selection of new sample candidates by using acquisition functions such as the expected improvement (EI) 
and (4) querying the new samples and updating the surrogate model. 
Repeat the above processes in (2) to (4) until the termination condition is met. 
\par 
The surrogate modeling and selection of new samples through acquisition function constitute the two key elements of EGO. 
Instead of purely using limited available high-fidelity (HF) samples to build a surrogate, 
a wide attention has been drawn to multi-fidelity surrogates (MFS), 
which augment the expensive HF samples by a large number of low-fidelity (LF) yet cheap samples to achieve better modeling accuracy. 
Generally, the MFS models can be categorized into three groups as 
(1) correlation function based modeling ~\cite{lewis_trust_1996,han_improving_2013},
(2) space mapping~\cite{bakr_an_2001, leifsson_multiobjective_2016} and 
(3) co-kriging methods~\cite{kennedy_predict_1998,forrester_multi-fidelity_2007, qian_bayesian_2008,park_low-fidelity_2018, rumpfkeil_multi-fidelity_2018}. 
In this paper, we focuse on using co-kriging to build a new multi-fidelity surrogate based optimization algorithm.  
\par 
On the other hand, when the selection of new sample candidates is concerned, 
we can use different acquisition functions such as EI and the probability of improvement (PI)~\cite{pickett_review_2011, liu_comparison_2012}. 
However, different from EGO algorithms that build on single-fidelity surrogate, 
we have to decide whether to use high- or low-fidelity simulations to query the new samples in MFS-based optimization (abbreviated as MFO).
Some pioneering works propose to only add the HF samples in the iteration process~\cite{hirschel_variable-fidelity_2010, zhou_active_2016}. 
By taking the cost ratio and the similarities between HF and LF models into account, some other works propose to adaptively infill both HF and LF samples in each optimization cycle~\cite{huang_sequential_2006, le_gratiet_recursive_2014, zhang_variable-fidelity_2018,hao_adaptive_2020}.  
\par 
Note that, most previous works as noted above only add in one sample per optimization cycle. 
However, as pointed by Ginsbourger et al. \cite{ginsbourger_multi-points_2007}, the parallelism is crucial to take advantage of the modern distributed systems to accelerate the optimization process. 
Specifically, when the simulations can be distributed over different processors, the wall-clock time (i.e., the time of total cycles rather than the times of total function calls) can be greatly reduced~\cite{viana2013efficient}. 
Meanwhile, to reduce the risks brought by unexpected simulation failure, 
querying multiple samples per optimization cycle is also encouraged in the engineering process.
Thus far, there are extensive studies on parallel optimization for single-fidelity surrogate based optimization~\cite{haftka2016parallel, zhan2017pseudo}. 
However, not much has been reported about paralleled optimization in MFO, 
where the key issue would be how to intelligently assign the high- and low-fidelity sample batches in each optimization cycle. 
\par 
To address the above sample assignment task in MFO, we propose a new infill criterion named as Filter-GEI. 
On one hand, in order to balance in between the local exploitation and global exploration~\cite{schonlau_computer_1997}, the GEI acquisition function will suggest new infill sample candidates that either close to the optima of the fitted surrogate or in the area of maximum prediction uncertainty. 
On the other hand, the LF model can more or less capture the global trend of the HF model over space when we decide to use an MFS to slove an optimization problem~\cite{giselle_fernandez-godino_issues_2019}. 
Therefore, we propose an adaptive filter function on top of the GEI, and we call the HF simulation only when the sample candidates have relatively better estimated objective value, while
we query the rest samples of GEI through LF simulation to explore the searching space globally. 
With the above, we can achieve a good balance in between local exploitation and global exploration by making full use of the multi-fidelity and parallel computing resources. 
\par 
\par 
The remainder of this paper is organized as follows: 
Section 2 presents a brief introduction of co-kriging and GEI acquisition function. 
Then, Section 3 details our proposed parallel infill criterion, Filter-GEI. 
After that, Section 4 shows the experimental studies on benchmark functions and an engineering problem. Finally, we draw conclusions in Section 5. 

\section{Preliminaries}

As the surrogate model and acquisition function are the two key elements of surrogate-based optimization, we introduce the basics of co-kriging and GEI acquisition function in this section.

\subsection{Co-kriging}

\par Co-kriging is a popularly used multi-fidelity surrogate technique, which models both the LF and HF functions as Gaussian process, of which mathematical expressions can be written as:

\begin{equation}
\begin{split}
\bm{Y}_{\text{LF}}(\bm{x})&=\mu_{\text{LF}}+{Z}_{\text{LF}}(\bm{x})\\
\bm{Y}_{\text{HF}}(\bm{x})&=\mu_{\text{HF}}+{Z}_{\text{HF}}(\bm{x})\\
\end{split}
\end{equation}
where $\mu_{\text{LF}}$ and $\mu_{\text{HF}}$ denote the mean function values of the LF and HF model; ${Z}_{\text{LF}}(x)$ and  ${Z}_{\text{HF}}(x)$ denote the random Gaussian variables. Then, the relation between the HF and LF functions are formulated as:
\begin{equation}
y_{\text{HF}}(\bm{x})=\rho \cdot y_{\text{LF}}(\bm{x})+{Z}_{\text{DF}}(\bm{x})
\end{equation}
where $\rho$ is the scaling factor, and $Z_{\text{DF}}(x)$ is the discrepancy function (DF) that models the differences between the HF function and the corrected LF function. 

For standard Gaussian process, the correlations between samples $\bm{x}_1$ and $\bm{x}_2$ can be described as:
\begin{equation}
Cov[{Z}(x_1),{Z}(x_2)]=\sigma^{2} 
exp(-\sum^{d}_{h=1} 
\theta_{h}||x_{1,h}-x_{2,h}||^{2})
\end{equation}

\noindent where $d$ denotes the dimension of the problem, 
$\sigma^2$ is the process variance, and $\theta_{h}$ represents the hyper-parameter which determines the bumpiness of fitted function curves. 
For co-kriging, it deals with Gaussian process of multiple fidelity models, the corresponding covariance matrix and correlation vector can be expressed as:

\begin{scriptsize}%
\begin{equation}
\bm{C}=\begin{pmatrix}

\sigma_{\text{LF}}^{2} R_{{\text{LF}}}(\bm{X}_{\text{LF}},\bm{X}_{\text{LF}})  \quad \rho\sigma_{\text{LF}}^{2} R_{{\text{LF}}}(\bm{X}_{\text{LF}},\bm{X}_{\text{HF}}) \vspace{0.2cm}\\

\rho\sigma_{\text{LF}}^{2} R_{{\text{LF}}}(\bm{X}_{\text{HF}},\bm{X}_{\text{LF}})  \quad \rho^{2}\sigma_{\text{LF}}^{2} R_{{\text{LF}}}(\bm{X}_{\text{HF}},\bm{X}_{\text{HF}})\vspace{0.05cm}\\
\qquad \qquad \qquad  \qquad  \qquad+\sigma_{\text{DF}}^{2} R_{{\text{DF}}}(\bm{X}_{\text{LF}},\bm{X}_{\text{LF}}) \\
\end{pmatrix}
\end{equation}
\end{scriptsize}

\begin{equation}
\bm{c} = \begin{pmatrix}
\rho\sigma_{\text{LF}}^{2}R_{{\text{LF}}}(\bm{X}_{\text{LF}},\bm{x}^*) \\
\rho^{2}\sigma_{\text{LF}}^{2}R_{{\text{LF}}}(\bm{X}_{\text{HF}},\bm{x}^*)
+\sigma_{\text{DF}}^{2}R_{{\text{DF}}}(\bm{X}_{\text{LF}},\bm{x}^*) \\
\end{pmatrix}
\end{equation}

\noindent where $\theta_{\text{LF}}$ and $\theta_{\text{DF}}$ are the bumpiness hyper-parameters of LF and DF models, respectively. 
Finally, the function prediction and corresponding prediction uncertainty of co-kriging model at a new point $\bm{x}^{*}$ are:  

\begin{small}
\begin{equation}
\begin{split}
\hat{y}_{\text{HF}}(\bm{x}^{*})=&\beta_{0}+\bm{c}^{T}\bm{C}^{-1}(y_{s}-\beta_{0}\bm{F})\\
\hat{s}_{\text{HF}}^{2}(\bm{x}^{*})=&\sigma^2 \{ 1-\bm{c}^{T}\bm{C}^{-1}c +(1-\bm{F}^{T}\bm{C}^{-1}\bm{c})^2/\bm{F}^{T}\bm{C}^{-1}\bm{F}    \}
\end{split}
\end{equation}
\end{small}
where $\bm{F} =   \bm{1}$ when the regression order is set as zero.

\subsection{Generalized expected improvement (GEI)}
Assuming $y({\bm{x}})$ comes from a Gaussian process with the posterior estimate as $y({\bm{x}}) \sim N(\hat y({\bm{x}}),{s ^2}({\bm{x}}))$, the basic idea of the EI acquisition function is to measure how much improvement of the objective function can be made with respect to the current best solution ${f_{\min }}$, which can be expressed as:
\begin{small}
\begin{equation}
\begin{split}
EI(\bm{x}) =& \begin{cases}
(f_{\text{min}}-\hat{y}(\bm{x}))\Phi(u(\bm{x}))+s(\bm{x})\phi(u(\bm{x}))  \quad (s  \,\textgreater\,\,0)\\
0  \qquad (s\leq 0)
\end{cases} \\
u(\bm{x})=&{(f_{\text{min}}-\hat{y}(\bm{x}))}/{s(\bm{x})}\\
\end{split}
\end{equation}
\end{small}where, $\Phi \left(  \cdot  \right)$ and $\phi \left(  \cdot  \right)$ denote the standard normal distribution and density functions.

Furthermore, Schonlau~\cite{schonlau_computer_1997} observed that, the classic EI acquisition function puts too much emphasis in the local search near the optima of the fitted surrogate, so he extended the classic EI to the generalized EI (GEI), of which formula is shown as below:
\begin{equation}
\begin{split}
E(I(\bm{x})^{g}) =s^{g}\sum_{k=0}^{g}(-1)^{k}
\begin{pmatrix} g \\ k  \end{pmatrix}
u^{g-k}T_{k}\\
 \begin{cases}
u={(f_{min}-\hat{y}(\bm{x}))}/{s(\bm{x})}\\
T_{0}=\Phi(u)\\
T_{1}=-\phi(u)\\
T_{k}=-u^{k-1}\phi{u}+(k-1)T_{k-2}
\end{cases}
\end{split}
\end{equation}
where the parameter $g$ can be set as $0,1,2 \cdots n$;
and $\begin{pmatrix} g \\ k  \end{pmatrix}$ denotes the combination. With the increase of $g$, the acquisition function will put more and more emphasize on the global search.
In particular, the GEI is degenerated to the classic EI when $g=1$. 

Note that, the GEI is proposed for single-fidelity surrogate based-optimization. However, for multi-fidelity surrogate based-optimization, a key issue would be how to adaptively distribute the new sample candidates for high- and low-fidelity simulations, respectively. To address the above sample assignment issue, we extend the GEI to Filter-GEI, which will be discussed in detail in Section 3.

\section{The Filter-GEI Method}
Different from single-fidelity surrogate based optimization, multi-fidelity surrogate based optimization augments the limited available high-fidelity samples by a large number of low-fidelity samples to boost the optimization performance, as shown in Eq.(9).
Then, a key factor that influences the effectiveness of multi-fidelity surrogate-based optimization is how to assign samples for high- or low-fidelity simulations in the iteration process.  

\begin{equation}
\begin{split}
&{\rm min}\quad y_{\text{HF}}(\bf{x})\\
&\bf{x} \in D\\
&{\rm with \quad assistance\quad of}\quad y_{\text{LF}}(\bf{x})
\end{split}
\end{equation}

To address the above sample assignment issue in MFO, we propose Filter-GEI, which adaptively assigns samples for HF and LF simulations in each optimization cycle.
In this section, we first illustrate the details of Filter-GEI, and then we use an 1D toy problem to present the scheme of Filter-GEI intuitively.

\begin{figure}[ht]
\begin{center}
\includegraphics[scale=0.8, trim = 0 0 0 0]{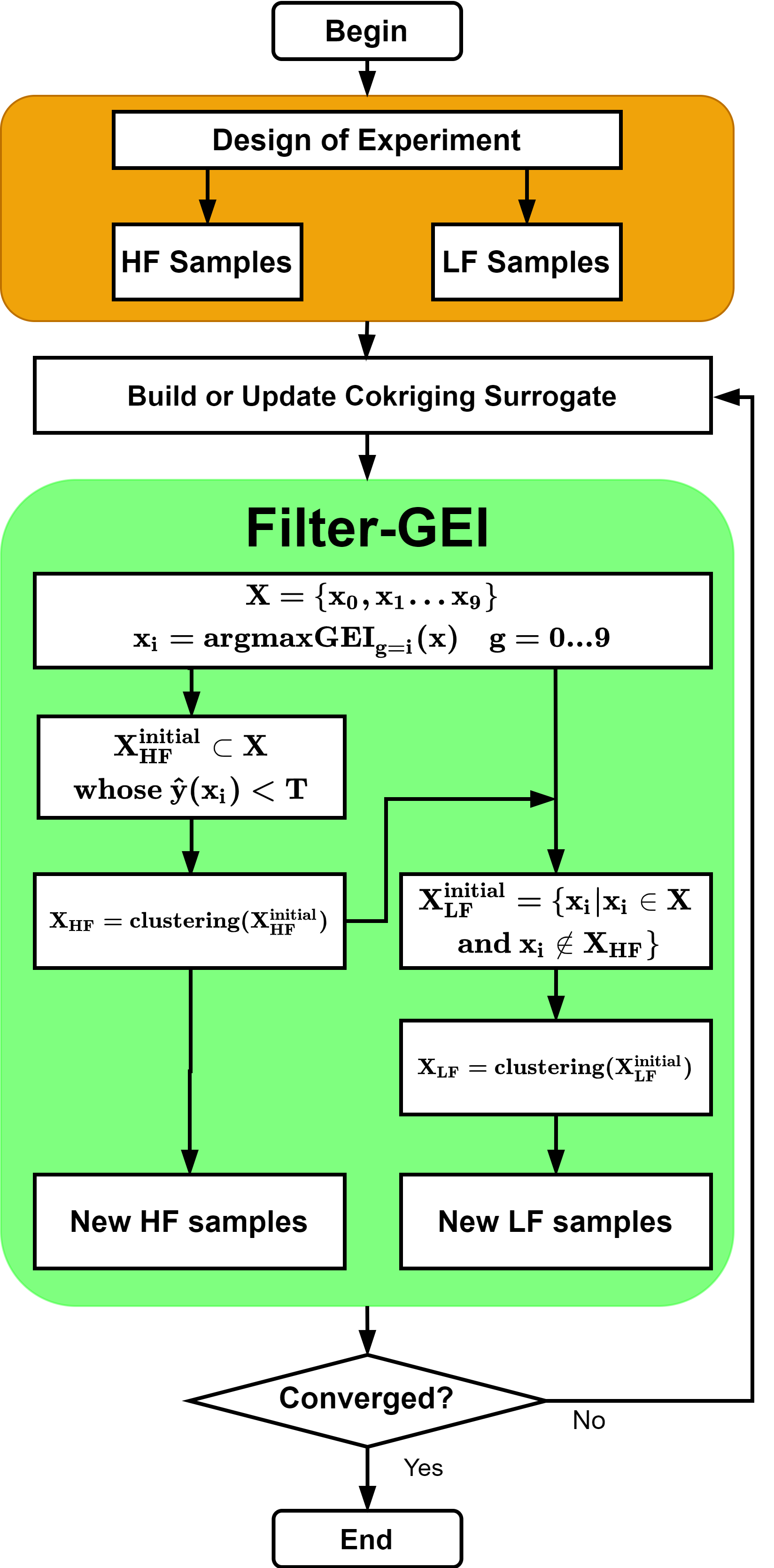}
\end{center}
\caption{Flowchart of the Filter-GEI Algorithm}   
\end{figure}

\subsection{Filter-GEI Acquisition Function}

The MFS is often used when we can only have a small number of expensive HF samples but plenty of cheap LF samples. 
More importantly, when we decide to use MFS in an engineering optimization problem, we usually have some prior knowledge that the LF model can more or less capture the global trend of the HF model, even if the real optimal solutions of HF and LF models may be not exactly overlapped in the design space. 

On the other hand, to take a balance in between local exploitation and global exploration in single-fidelity surrogate based optimization, the GEI acquisition function~\cite{schonlau_computer_1997} often suggests a set of sample candidates either in the neighborhood of the optima of the fitted surrogate model (exploitation) or in the area of of maximum prediction uncertainty (exploration).

With the above facts in mind, we propose Filter-GEI, which calls the expensive HF simulations only when the new sample candidates have relatively better estimated objective function value, while the rest sample batches of GEI will be evaluated through LF simulation. 
In other words, we propose to impose a filter over GEI, thereby we can query HF samples to be more focused on exploiting the "promising" areas with relatively better objective function values. In the meantime, we query the other sample candidates of GEI through LF simulations to explore the optimization space globally and reduce the model uncertainty. This way we can make full use of the multi-fidelity and parallel computing resources to balance in between local and global search. 

\subsubsection{Formulation of the filter over GEI}
To filter HF samples from the original GEI sample candidates, we propose the following threshold expression in Filter-GEI:
\begin{equation}
\begin{split}
T = & \omega \cdot  y_{\text{HF-min}}+(1-\omega) \cdot y_{\text{HF-mean}}\\
\end{split}
\end{equation}
where, $y_{\text{HF-min}}$ is the current best optimal solution, $y_{\text{HF-mean}}$ is a weighted averaged function value over all training samples, and $\omega$ is an adaptive weight coefficient that weighs the importance of $y_{\text{HF-min}}$ and $y_{\text{HF-mean}}$ in determining the threshold $T$.

Furthermore, by taking the correlations between HF and LF models into account, we propose a dynamic weight to adjust the importance of $y_{\text{HF-min}}$ and $y_{\text{HF-mean}}$ as below:
\begin{equation}
\begin{split}
\omega =& \frac{1}{1+(\hat{\sigma}_{\text{DF}}^{2}/\hat{\sigma}_{\text{LF}}^{2})^{0.5}}
\end{split}
\end{equation}
where, $\sigma_{\text{DF}}^2$ and  $\sigma_{\text{LF}}^2$ are the Gaussian process variance of DF and LF functions. 

When the HF and LF models are highly correlated, the value of $\hat{\sigma}_{\text{DF}}/\hat{\sigma}_{\text{LF}}$ should be small, and the corresponding co-kriging surrogate can be expected to be in good accuracy. 
Then, the weight of $y_{\text{HF-min}}$ will becomes much higher. In other words, the Filter-GEI will make the HF simulations to be even more focused on the local exploitation of "promising" areas with relatively better estimated objective function values.
On the contrary, when the correlations HF and LF models are not high, we will add more weight on $y_{\text{HF-mean}}$. In other words, the Filter-GEI will query HF samples in a much larger area to avoid the misleading of LF model and enhance the global exploration as well.

In addition, in order to take the convergence history into account, we propose a weighted averaged function value over all samples as shown in Eq.(12):
\begin{equation}
y_{\text{HF-mean}} = \frac{1}{{iter}}\sum^{{iter}}_{i = 0}\left( \frac{1}{n_{i}}\sum^{n_{i}}_{j=1}y_{HF-ij}\right)
\end{equation}
where, $iter$ denotes the number of total iterations thus far, $n_{i}$  is the number of newly added HF samples in the $i^{th}$ iteration, and $y_{HF-ij}$ denotes one of the HF sample that was added in the $i^{th}$ iteration.
From Eq.(12), we can see the weights of the samples that added in the same iteration are the same, while the sample weights of different iterations varies. More importantly, the sample weights are proportional to the number of newly added samples in a iteration. Such treatment is helpful to smooth the history curve of $T$ in Eq.(10). 

With the above, the filter function we proposed in Eq.(10) can be adaptively changed, which helps to take a balance in between global and local search when assigning the HF and LF samples.

\begin{algorithm}  
\caption{Filter-GEI based MFO Algorithm}  
\begin{algorithmic}[1]
\Require MF surrogate(contained all parameters), $n$  
\Ensure $X_{\text{HF}}$(the HF sample set); $X_{\text{LF}}$(the LF sample set) 
\Function {gei\_search}{$\hat{Y}_{m}$}  
\State $n \gets 10$
\For{$i= 1 \to n$} 
\State $g \gets i-1$
\State $[Xc_{i},Yc_{i}]\gets argmax[E(I^{g})]$
\EndFor
\State \Return{$[Xc_{i},Yc_{i}] \quad i=1\cdots n$}  
\EndFunction  
\State  
\Function{get\_threshold}{$\hat{Y}_{m}$}  
\State $ratio \gets \hat{\sigma}_{d}/\hat{\sigma}_{\text{LF}}$ 
\State $\omega \gets 1/(1+ratio)$  
\State $y_{min} \gets min[y_{sample}]$ 
\State $y_{mean} \gets mean[y_{sample}]$  
\State $T \gets \omega \; y_{h-min}+(1-\omega)y_{h-mean}$  
\State \Return{$T$}  
\EndFunction  
\State 
\Function {cluster}{$X,term$}  
\State  $[a,b] \gets argmin [dist(x_{i},x_{j})]$
\State  $dist_{min}\gets dist(x_{a},x_{b})$
\While $\quad dist_{min}<term$
\State  remove point in $x_{a}$ and $x_{b}$ \\ \qquad \quad who is closer to the existing samples 
\State  $[a,b] \gets argmin [dist(x_{i},x_{j})]$ $x_{i},x_{j} \in X$
\State  $dist_{min}\gets dist(x_{a},x_{b})$
\EndWhile
\State \Return{$X_{left}$}  
\EndFunction  
\State 
\Function{filter}{$\hat{Y}_{m}$} 
\State $[Xc_{i},Yc_{i}] \gets GEI\_search\{\hat{Y}_{m}\}$
\State$T \gets get\_threshold\{\hat{Y}_{m}\}$
\State $j \gets 0$ 
\State $k \gets 0$ 
\For{$i= 1 \to n$}
\If{$Yc_{i}<T$}
\State $j \gets j+1$
\State $[Xh_{j},Yh_{j}] \gets[Xc_{i},Yc_{i}]$
\Else 
\State $k \gets k+1$
\State $[Xl_{k},Yl_{k}] \gets[Xc_{i},Yc_{i}]$
\EndIf
\EndFor
\State $X_{\text{HF}} \gets  cluster\{X_{\text{HF}}^{initial},d^T_{\text{HF}}\}$
\State $X_{\text{LF}}^{initial} \gets \{x|x \in X_{c_i} \text{ and } x \notin X_{\text{HF}}\}$
\State $X_{\text{LF}}\gets cluster\{ X_{\text{LF}}^{initial},d^T_{\text{LF}} \}$
\State \Return{$X_{\text{HF}},X_{\text{LF}}$}  
\EndFunction       
\end{algorithmic}  
\end{algorithm}  
\subsubsection{Sample clustering}

Note that, when using the GEI acquisition function to select new sample candidates, the suggested samples of GEI with different g values may get overlapped, resulting in ill-conditioning of the covariance matrix. To avoid the above issue and thus ensure the normal implementation of Filter-GEI based MFO, we use hierarchical clustering like the work in \cite{ponweiser_clustered_2008}. The steps of hierarchical clustering are listed as below:

(1) Treat each sample candidate as a independent class and calculate the minimum distance between classes;

(2) Combine the two classes with the smallest distance into a new class;

(3) Recalculate the distance between the new class and all classes;

(4) Repeat (2) and (3) until the minimum distance between classes is larger than some value $d^T$.

Furthermore, by taking the cost of HF and LF samples into account, we use the following termination conditions when clustering samples for HF and LF, respectively:

\begin{equation}
\begin{split}
d^T_{\text{HF}} &= 0.1\sqrt{d} \\
d^T_{\text{LF}} &= d^T_{\text{HF}}\left(\frac{cost_{\text{LF}}}{cost_{\text{HF}}}\right)^{\frac{1}{3}}
\end{split}
\end{equation}

\noindent where $d$ denotes the dimension of the problem.

Combining the filter in Eq.(10) and sample clustering, the procedure of Filter-GEI is carried out as follows:
After obtaining the sample candidates using the GEI in an optimization cycle, we first select the HF samples through Eq.(10), and we control the number of new HF samples to query through sample clustering. After that, we select the LF samples from the rest of GEI sample candidates.
Based on Filter-GEI, we propose a new MFO algorithm as shown in Fig.1. The related pseudo code is also presented in Algorithm 1.

\subsection{An illustrative example}

In this subsection, we use an  one-dimensional toy problem to show the scheme of Filter-GEI based MFO intuitively. 
Equation (14) shows the mathematical expressions of the toy problem, and we carry out optimization for this toy function as a minimization problem.

\begin{equation}
\begin{split}
y_{\text{HF}}=&(6\bm{x} -2)^{2}sin(12\bm{x}-4)\\
y_{\text{LF}}=&0.5y_{\text{HF}}+10(\bm{x}-0.5)-5\\
&\bm{x} \in [0,1]
\end{split}
\end{equation}

Initially, we use 3 HF samples and 5 LF samples to build a co-kriging surrogate, and then we use the Filter-GEI algorithm to carry out the optimization, as shown in Fig.2. Table 1 lists the sample details in each optimization cycle. 

In Fig.2, the hollow squares denote the original sample candidates obtained from GEI, while the hollow circles denote the new samples to be queried through LF simulation, the threshold to filter HF samples is represented by dashed straight line. 
As shown in Fig.2(a), the sample candidates originally suggested by GEI are mainly gathered around $x = 0.26$ and $x=0.70$. 
Through the filter using Eq.(10) and sample clustering as well, the sample x=0.7053 is selected for HF simulation in this iteration. Thereafter, we cluster the rest GEI sample candidates and select x = 0.6504 and x=0.2562 to be evaluated through LF simulation in this optimization cycle.

Similarly, in the second iteration (see Fig.2(b)), the Filter-GEI assigns the boundary sample candidate as LF samples to reduce the uncertainty of co-kriging, while the HF simulation is called to exploit the "promising" area with relatively better estimated objective function value. 
Finally, as shown in Fig.2(d), the Filter-GEI successfully finds the real optimal solution of the toy problem. In the meantime, the HF function is perfectly fitted by the co-kriging, even though no samples locate at the boundaries of the HF curve.

With the above, the motivation of Filter-GEI is well demonstrated. That is, while balancing in between the local exploitation and global exploration, we can filter HF samples to be more focused on exploiting "promising" areas with relatively better objective function value. In the meantime, considering the correlations between HF and LF models, we can query more LF samples in the rest of areas to reduce the model uncertainty over space. This way we can make full use of multi-fidelity and parallel computing resources to achieve the optimal solution most efficiently and effectively. 

\begin{figure*}[htbp]
\centering
\subfigure[iteration 1]{
\begin{minipage}[t]{0.45\linewidth}
\centering
\includegraphics[scale=0.6]{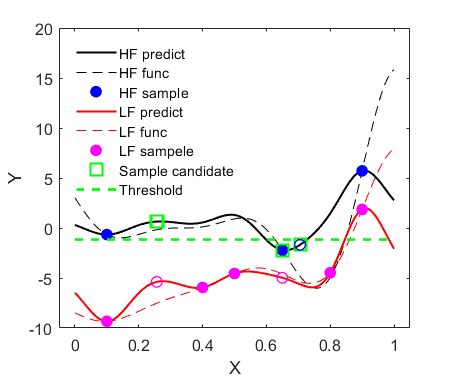}
\end{minipage}%
}%
\subfigure[iteration 2]{
\begin{minipage}[t]{0.45\linewidth}
\centering
\includegraphics[scale=0.6]{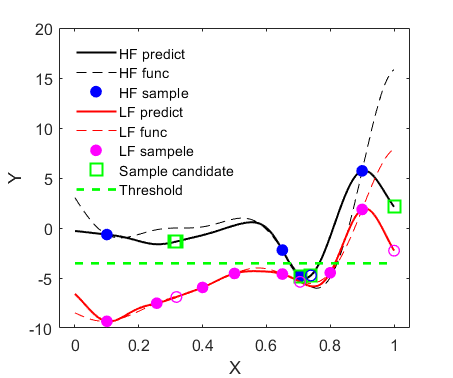}
\end{minipage}
}%

\subfigure[iteration 3]{
\begin{minipage}[t]{0.45\linewidth}
\centering
\includegraphics[scale=0.6]{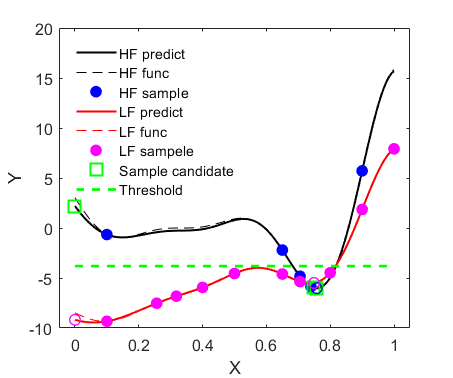}
\end{minipage}
}%
\subfigure[well fitted]{
\begin{minipage}[t]{0.45\linewidth}
\centering
\includegraphics[scale=0.6]{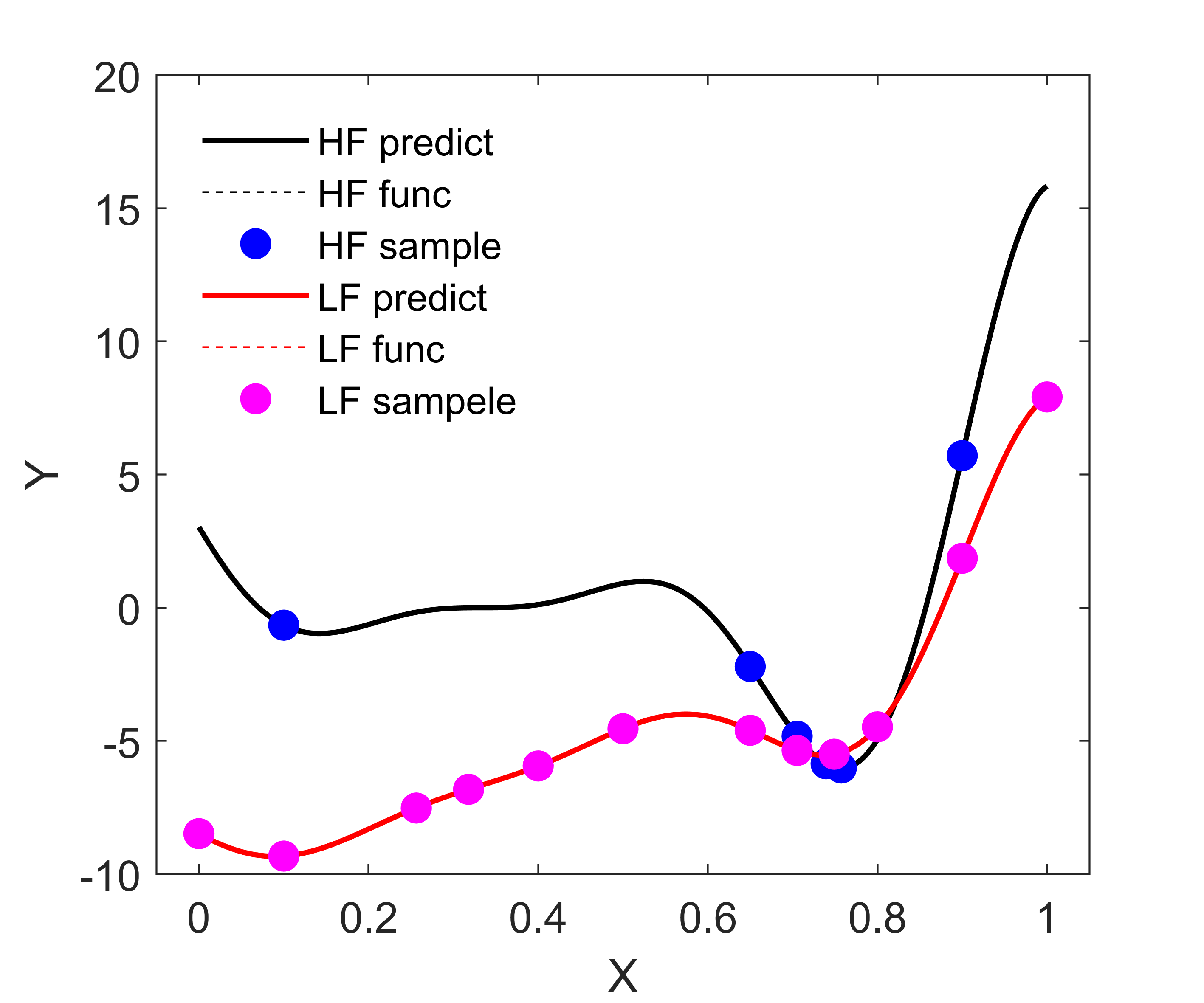}
\end{minipage}
}%

\centering
\caption{The optimization process of the one-dimensional toy  function by using Filter-GEI}
\end{figure*}

\begin{table}[ht]
\centering
\caption{The sampling detail of illustrative optimization example}
\renewcommand\arraystretch{1.2}
\begin{tabular}{cccc}
\hline
    Iteration &\tabincell{c}{ HF points} & \tabincell{c}{LF points} & T \\
\hline
    initial & [0.1;0.65;0.9] & [0.1;0.4;0.5;0.8;0.9] & \textbackslash{} \\
    1     & 0.7053 & [0.2562;0.6504] & -1.160 \\
    2     &0.7395 & [0.3180;0.7053;1.0000] & -3.529 \\
    3     & 0.7574 &[0.0000;0.7490] & -3.808 \\
\hline
    \end{tabular}
\end{table}
\par%

\section{Experimental tests}
In this section, we test the proposed Filter-GEI algorithm and compare it with a set of the state of the art MFO algorithms on both benchmark functions and an engineering optimization design problem. 

\subsection{Baseline for comparison and experimental setup}
To verify the effectiveness of Filter-GEI, we compare it with 
two the state of the art algorithms as augmented-EI\cite{huang_sequential_2006} and VF-EI\cite{zhang_variable-fidelity_2018}. 
Specifically, the augmented-EI is the first MFO algorithm which queries both HF and LF samples in the iteration process. It takes both the similarities and the cost ratio between HF and LF samples into account when deciding whether to call the HF or LF simulation. 
On the other hand, VF-EI is a recently proposed MFO algorithm, which takes a special consideration on the model uncertainty to improve the performance of MFO.

Note that, both augmented-EI and VF-EI add one sample in each optimization cycle. Differently, our proposed Filter-GEI can be first parallel MFO algorithm, which adaptively assign sample batches for HF and LF simulation in each iteration.
The comparisons between Filter-GEI and augmented-EI and VF-EI shall manifest the advantage of parallel optimization in accelerating the process. 

To show the role of filtering of HF samples, we also compare Filter-GEI with the GEI-based MFO, which directly add the sample batches of GEI for both HF and LF simulations in each iteration. In addition, we also test EI-based MFO, which add in one sample for both HF and LF evaluations per optimization cycle. 
Note that, infilling both LF and HF samples at the same location is helpful to improve the accuracy of discrepancy function, though it cannot provide more information of HF model in the same location.

For fair comparison, we use the open source code ooDACE~\cite{couckuyt2014oodace} to build a uniform test bed for all the comparing algorithms. 
For each testing case, the number of initial HF and LF samples is set as three and six times of the function dimension, respectively. We use the strategy of nearest neighbor sampling~\cite{guo_analysis_2018} to generate these HF and LF samples.

\begin{table*}[htbp]
  \centering
  \caption{Definition of multi-fidelity test cases}
  \renewcommand\arraystretch{1.2}
    \begin{tabular}{cllcc}
    \toprule
    No.    & HF function & LF function & Dimension & Termination error\\
    \midrule
    case1 & Forrester1 & Forrester1a & 1 & 0.0602 (1\%)\\
    case2 & Forrester1 & Forrester1b & 1  & 0.0602 (1\%)\\
    case3 & Hartman3 & Hartman3+MA3*7.6 & 3 &0.0386 (1\%)\\
    case4 & Ackley5 & Ackley5+MA5*0.74 & 5 & 0.1000 (optimal value is 0)\\
    case5 & Hartmann6 & Hartmann6+MA6*0.21 & 6 & 0.0304 (1\%)\\
    \bottomrule
    \end{tabular}%
  \label{tab:addlabel}%
\end{table*}%

\begin{table*}[htbp]
  \centering
  \caption{Description of test function}
    \begin{tabular}{ll}
    \toprule
    Name  & \multicolumn{1}{c}{Description} \\ 
    \midrule
    Forrester1 & $\begin{array}{l}
{f}(x) = {(6x - 2)^2}\sin (12x - 4)\\  
x \in [0,1]{\rm{     }},\quad{x^*} = 0.7525,f* =  - 6.020740
\end{array}$ \\   \vspace{0.5cm}
    Forrester1a & ${f}(x) = 0.5{f}(x) + 10(x - 0.5) - 5$ \\  \vspace{0.5cm}
    Forrester1b & ${f}(x) = {f}(x) - 5$ \\  \vspace{0.5cm}
    Hartman3 & $\begin{array}{l}
{f}(x) =  - \sum\limits_{i = 1}^4 {{c_i}\exp [ - \sum\limits_{j = 1}^3 {{\alpha _{ij}}{{({x_j} - {p_{ij}})}^2}} ]} \\
{\rm{where  }}\quad \alpha  = \left[ {\begin{array}{*{20}{c}}
3&{10}&{30}\\
{0.1}&{10}&{35}\\
3&{10}&{30}\\
{0.1}&{10}&{35}
\end{array}} \right]{\rm{    }}c = \left[ {\begin{array}{*{20}{c}}
1\\
{1.2}\\
3\\
{3.2}
\end{array}} \right]{\rm{   }}p = \left[ {\begin{array}{*{20}{c}}
{0.3689}&{0.1170}&{0.2673}\\
{0.4699}&{0.4387}&{0.7470}\\
{0.1091}&{0.8732}&{0.5547}\\
{0.0381}&{0.5743}&{0.8828}
\end{array}} \right]\\
x \in [0,1]{\rm{       }},\quad{x^*} = (0.114,0.556,0.852),\quad{f^*} =  - 3.8627
\end{array}$  \\  \vspace{0.5cm}
    MA3   & $\begin{array}{l}
f(x) = 0.585 - 0.324{x_1} - 0.379{x_2} - 0.431{x_3} - 0.208{x_{\rm{1}}}{x_2} + 0.326{x_1}{x_3}\\
 + 0.193{x_2}{x_3} + 0.225{x_1}^2 + 0.263{x_2}^2{\rm{  + }}0.274{x_3}^2
\end{array}$ \\  \vspace{0.5cm}
    Ackley5 &  $\begin{array}{l}
{f}(x) =  - a\exp \left[ { - b\sqrt {\frac{1}{n}\sum\limits_{i = 1}^5 {{x_i}^2} } } \right] - \exp \left[ {\frac{1}{n}\sum\limits_{i = 1}^5 {\cos (c{x_i})} } \right] + a + \exp (1)\\
a = 20;b = 0.2;c = 2\pi \\
{x_i} \in [ - 2,2]{\rm{          }},\quad{x^*} = (0,0,0,0,0),\quad{f^*} = 0
\end{array}$\\  \vspace{0.5cm}
    MA5   & $\begin{array}{l}
f(x) = 0.585 - 0.00127{x_1} + 0.00113{x_2} + 0.00663{x_3} + 0.0129{x_4} + 0.00611{x_5}\\
 + 0.00526{x_1}{x_4} + 0.0106{x_1}{x_5} - 0.000626{x_2}{x_4} - 0.00310{x_2}{x_5}\\
 - 0.00724{x_4}{x_5} + 0.00096{x_3}^2 + 0.0124{x_4}^2 + 0.0101{x_5}^2
\end{array}$ \\  \vspace{0.5cm}
    Hartmann6 & $\begin{array}{l}
{f}(x) =  - \sum\limits_{i = 1}^4 {{c_i}\exp [ - \sum\limits_{j = 1}^3 {{\alpha _{ij}}{{({x_j} - {p_{ij}})}^2}} ]} \\  \vspace{0.5cm}
{\rm{where  }}\quad \alpha  = \left[ {\begin{array}{*{20}{c}}
{{\rm{10}}}&{\rm{3}}&{{\rm{17}}}&{{\rm{3}}{\rm{.50}}}&{{\rm{1}}{\rm{.7}}}&{\rm{8}}\\
{{\rm{0}}{\rm{.05}}}&{{\rm{10}}}&{{\rm{17}}}&{{\rm{0}}{\rm{.1}}}&{\rm{8}}&{{\rm{14}}}\\
{\rm{3}}&{{\rm{3}}{\rm{.5}}}&{{\rm{1}}{\rm{.7}}}&{{\rm{10}}}&{{\rm{17}}}&{\rm{8}}\\
{{\rm{17}}}&{\rm{8}}&{{\rm{0}}{\rm{.05}}}&{{\rm{10}}}&{{\rm{0}}{\rm{.1}}}&{{\rm{14}}}
\end{array}} \right]{\rm{    }}c = \left[ {\begin{array}{*{20}{c}}
1\\
{1.2}\\
3\\
{3.2}
\end{array}} \right]{\rm{   }}\\
p = {\rm{1}}{{\rm{0}}^{{\rm{ - 4}}}}\left[ {\begin{array}{*{20}{c}}
{{\rm{1312}}}&{{\rm{1696}}}&{{\rm{5569}}}&{{\rm{124}}}&{{\rm{8283}}}&{{\rm{5886}}}\\
{{\rm{2329}}}&{{\rm{4135}}}&{{\rm{8307}}}&{{\rm{3736}}}&{{\rm{1004}}}&{{\rm{9991}}}\\
{{\rm{2348}}}&{{\rm{1451}}}&{{\rm{3522}}}&{{\rm{2883}}}&{{\rm{3047}}}&{{\rm{6650}}}\\
{{\rm{4047}}}&{{\rm{8828}}}&{{\rm{8732}}}&{{\rm{5743}}}&{{\rm{1091}}}&{{\rm{381}}}
\end{array}} \right]\\
x \in [0,1]{\rm{    }},{x^*} = (0.20169,0.150011,0.476874,0.275332,0.311652,0.6573),{f^*} =  - 3.04246
\end{array}$ \\
    MA6   & $\begin{array}{l}
f(x) = 0.5{\rm{44 + }}0.{\rm{15984}}{x_1} + 0.{\rm{18857}}{x_2} + 0.0{\rm{2371}}{x_3} - 0.04144{x_4} + 0.3313{x_5}\\
 - 0.11813{x_6} - 0.03719{x_1}{x_2} + 0.03099{x_2}{x_3} - 0.0504{x_3}{x_6} - 0.03103{x_4}{x_5}\\
 + 0.19345{x_1}{x_6} - 0.2635{x_1}^2 - 0.17694{x_2}^2 - 0.23528{x_5}^2
\end{array}$\\
\cmidrule{1-2}   
\end{tabular}
\end{table*}

\begin{table*}[htbp]
\centering
\caption{Optimization results of Filter-GEI, augmented-EI, EI, GEI, and VF-EI on five benchmark functions}
\renewcommand\arraystretch{1.2}
\begin{tabular}{clccccc}
 \toprule
Function&  Algorithms & Iteration & Optimal &HF& LF &\tabincell{c}{Success\\rate} \\
\hline
     case1 & Filter-GEI &   $2\pm1$ & $-6.0103\pm0.0146$ &   $6\pm1$ &  $11\pm2$ &      20/20 \\
           
           &        GEI &   $3\pm1$ & $-6.0111\pm0.0171$ &  $10\pm1$ &  $11\pm3$ &      20/20 \\

           &         EI &   $3\pm1$ & $-6.0104\pm0.0165$ &   $6\pm1$ &   $9\pm1$ &      20/20 \\

           &      VF-EI &   $5\pm2$ & $-6.0150\pm0.0150$ &   $6\pm2$ &   $8\pm1$ &      20/20 \\

           & augmented-EI &   $5\pm3$ & $-6.0117\pm0.0115$ &   $6\pm2$ &   $8\pm1$ &      20/20 \\
\cline{2-7}
     case2

           & Filter-GEI &   $3\pm1$ & $-6.0115\pm0.0142$ &   $6\pm1$ &  $11\pm2$ &      20/20 \\
           
            &        GEI &   $3\pm1$ & $-6.0044\pm0.0199$ &  $11\pm3$ &  $11\pm4$ &      20/20 \\

           &         EI &   $3\pm1$ & $-6.0082\pm0.0136$ &   $6\pm1$ &   $9\pm1$ &      20/20 \\

           &      VF-EI &   $5\pm2$ & $-6.0191\pm0.0058$ &   $5\pm2$ &   $9\pm1$ &      20/20 \\

           & augmented-EI &   $3\pm1$ & $-6.0114\pm0.0131$ &   $5\pm1$ &   $7\pm1$ &      20/20 \\

\cline{2-7}
     case3       

           & Filter-GEI &   $4\pm1$ & $-3.8450\pm0.0107$ &  $16\pm2$ &  $39\pm5$ &      20/20 \\
           
            &  GEI &   $4\pm1$ & $-3.8491\pm0.0093$ &  $27\pm4$ &  $32\pm9$ &      20/20 \\

           &         EI &   $9\pm4$ & $-3.8457\pm0.0101$ &  $18\pm4$ &  $27\pm4$ &      20/20 \\

           &      VF-EI &  $13\pm5$ & $-3.8507\pm0.0099$ &  $16\pm4$ &  $24\pm4$ &      20/20 \\

           & augmented-EI &  $10\pm3$ & $-3.8511\pm0.0103$ &  $15\pm2$ &  $21\pm2$ &      20/20 \\
\cline{2-7}
     case4 

           & Filter-GEI &  $20\pm3$ & $0.0770\pm0.0164$ &  $37\pm3$ & $92\pm13$ &      20/20 \\
           
           &        GEI &  $16\pm4$ & $0.0744\pm0.0178$ &  $64\pm4$ & $76\pm12$ &      20/20 \\

           &         EI &  $25\pm6$ & $0.0713\pm0.0183$ &  $40\pm6$ &  $55\pm6$ &      20/20 \\

           &      VF-EI & $47\pm24$ & $0.4945\pm0.9972$ &  $16\pm0$ & $76\pm24$ &      12/20 \\

           & augmented-EI & $30\pm14$ & $0.0673\pm0.0223$ &  $28\pm5$ & $47\pm11$ &      20/20 \\
\cline{2-7}
     case5

           & Filter-GEI &  $12\pm4$ & $-3.0312\pm0.0081$ &  $33\pm6$ & $94\pm24$ &      20/20 \\
           
            &        GEI &  $17\pm9$ & $-3.0115\pm0.0215$ & $123\pm91$ & $141\pm101$ &      14/20 \\

           &         EI & $25\pm22$ & $-3.0176\pm0.0147$ & $43\pm22$ & $61\pm22$ &      18/20 \\

           &      VF-EI & $42\pm30$ & $-3.0216\pm0.0263$ &  $23\pm4$ & $73\pm29$ &      15/20 \\

           & augmented-EI & $51\pm33$ & $-3.0058\pm0.0231$ &  $36\pm8$ & $69\pm30$ &      12/20 \\
\hline
\end{tabular}  
\end{table*}
\par

\subsection{Tests and discussions on benchmark functions}
We use the benchmark functions in \cite{zhang_variable-fidelity_2018} to carry out the optimization tests, the definitions and descriptions of the benchmark functions are shown in Tables 2 and 3.
As the real optimal solution of benchmark functions are known, the
comparing algorithms shall be terminated when the relative error between the obtained optimal solution and the real one less than 1\% or the number of iterations is 15 times of the function dimension. 
As an exception, since the optimal solution of the Case 4 in Table 2 is 0, we follow the treatment in \cite{zhang_variable-fidelity_2018}, i.e., the algorithms will be stopped once the absolute error between the obtained optimal solution and the real one was less than 0.1. 
In addition, each benchmark function will run 20 times, the mean and standard deviation of the obtained optimal solutions will be calculated to compare the performance of different algorithms.

Tables 4 shows the comparing results of the five benchmark functions.
When comparing between Filter-GEI with GEI- and EI-based MFO, the Filter GEI always uses the minimum number of HF samples to achieve the competitive optimal solution. 
As a contrast, without filtering of HF samples, the GEI-based MFO need significant more HF samples. 
In the meantime, as the Filter-GEI and GEI-based MFO add multiple samples instead of one sample in each iteration, these two algorithms usually need much less iterations than EI-based MFO to meet the termination condition. 
Meanwhile, the Filter-GEI consumes the same or even less HF samples than EI-based MFO to achieve the competitive results. 
To summarize, the Filter-GEI outperforms GEI-based MFO in terms of the number of HF samples, and it has better efficiency than EI-based MFO in terms of both iterations and the number of HF samples. 

When comparing to the VF-EI and augmented-EI, Filter-GEI needs significant fewer iterations to achieve comparable optimal solution. 
More importantly, as shown in the last column of Table 4, the VF-EI
only succeed to achieve the optimal solution 12 times out of 20 tests for the five-dimensional Ackley5 function (case 4), and the augmented-EI has a success rate of 12/20 for the six-dimensional Hartman6 function (case 5).
On the contrary, Filter-GEI succeed to find the optimal solution for all the testing cases. 
In the meantime, as shown in the first column of Table 4, the fluctuations of iterations of Filter-GEI is much smaller than that of VF-EI and augmented-EI. 
The above results indicate that, in addition to having significantly better efficiency in terms of wall-clock time (i.e., the number of iterations rather than the number of simulations), the performance of Filter-GEI is also much more robust than those of VF-EI and augmented-EI. 
With the above, the effectiveness of Filter-GEI and its advantage over other the state of the art algorithms have been well demonstrated. 

\subsection{Engineering case study}

In this subsection, we test Filter-GEI on a turbine design problem to further validate its effectiveness. 

\subsubsection{Problem description}
We select the first-stage stator vane of the well-known Energy Efficient Engine (GE-$\text{E}^3$) \cite{cherry_aerodynamic_1984} turbo for the test. The goal is to minimize the energy loss of the stator vane through optimizing the contour of stator vane.

We use the Bezier curves to parameterize the stator vane as shown in Fig.3. The profile is composed of three Bezier curves and two arcs. Each Bezier curve is adjusted by four control points, and the variation of the 12 control points are determined by 18 design parameters.
In this study, we select seven variables from the 18 design parameters to optimize the stator profile. The definition and variable range of these seven variables are shown in Table 5.

\begin{figure}[ht]
\begin{center}
\includegraphics[scale=0.79]{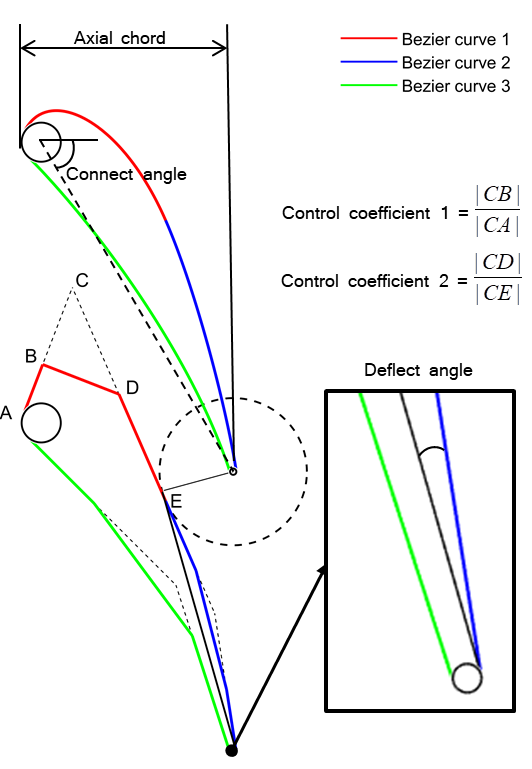}
\end{center}
\caption{Parameterization method for the engineering design}
\label{fig:1-1}      
\end{figure}
\begin{table}[h]
\caption{Design variables in the engineering design case}
\begin{center}
\renewcommand\arraystretch{1.2}
\begin{tabular}{cccc}
\hline
  \tabincell{c}{No.} &  \tabincell{c}{Geometric\\ definition} & \tabincell{c}{Reference\\value }& \tabincell{c}{Variation\\ range} \\
\hline
        1 & axial chord &      33.9[$^\circ$]  &     (-2.0, 2.0) \\
        2 &center connect angle &       59.8[$^\circ$]  &     (-4.0, 2.0) \\
        3 &\tabincell{c}{inlet upper wedge\\ angle} &         69.0[$^\circ$] &    (-15.0, 3.0) \\
        4 & outlet deflect angle &        4.5[$^\circ$]  & (-1.5, 4.5) \\
        5 & correlation coefficient &       0.35[-]  & (-0.05, 0.10) \\
        6 & control coefficient 1 &        0.40[-] & (-0.15, 0.15) \\
        7 & control coefficient 2 &        0.50[-]  & (-0.15, 0.15) \\
\hline
\end{tabular}  
\end{center}
\end{table}

\subsubsection{High- and Low-Fidelity Simulation Models}
The definition of the energy loss is shown in Eq.(15):
\begin{equation}
\xi  =(1-\frac{h_{1-1}+0.5c_{1-1}^{2}}{h_{0-0}+0.5c_{0-0}^{2}})\times 100\%
\end{equation}
where $h$ denotes the enthalpy, $c$ denotes the flow velocity at different locations of stator vane. The subscripts $0-0$ and $1-1$ denote the inlet and outlet of the stator, respectively.

To obtain the energy loss of stator vane, we use the commericial software ANSYS CFX 14.5 to carry out the HF and LF simulations. 
Specifically, we employ the Reynolds Averaged Navier-Stokes Equations (RANS) coupled by the Spalart-Allmaras turbulence model for the simulation process.
The boundary conditions at the inlet and outlet of the stator is shown in Table 6, by referring to the descriptions of GE-$\text{E}^3$ turbine in \cite{cherry_aerodynamic_1984}.
The grid of stator is automatically generated using one O-grid block and three H-grid blocks by using an in-house code, and Fig.4 presents the test on the grid independence.

Similar to~\cite{zhang_variable-fidelity_2018}, the HF and LF simulation models are obtained by using different cell number. In this study, the cell number in HF and LF models are set as 100701 and 12141, respectively. Correspondingly, the cost ratio of LF and HF models is 0.1. 

\begin{figure}[ht]
\begin{center}
\includegraphics[scale=0.58,trim=55 0 0 0]{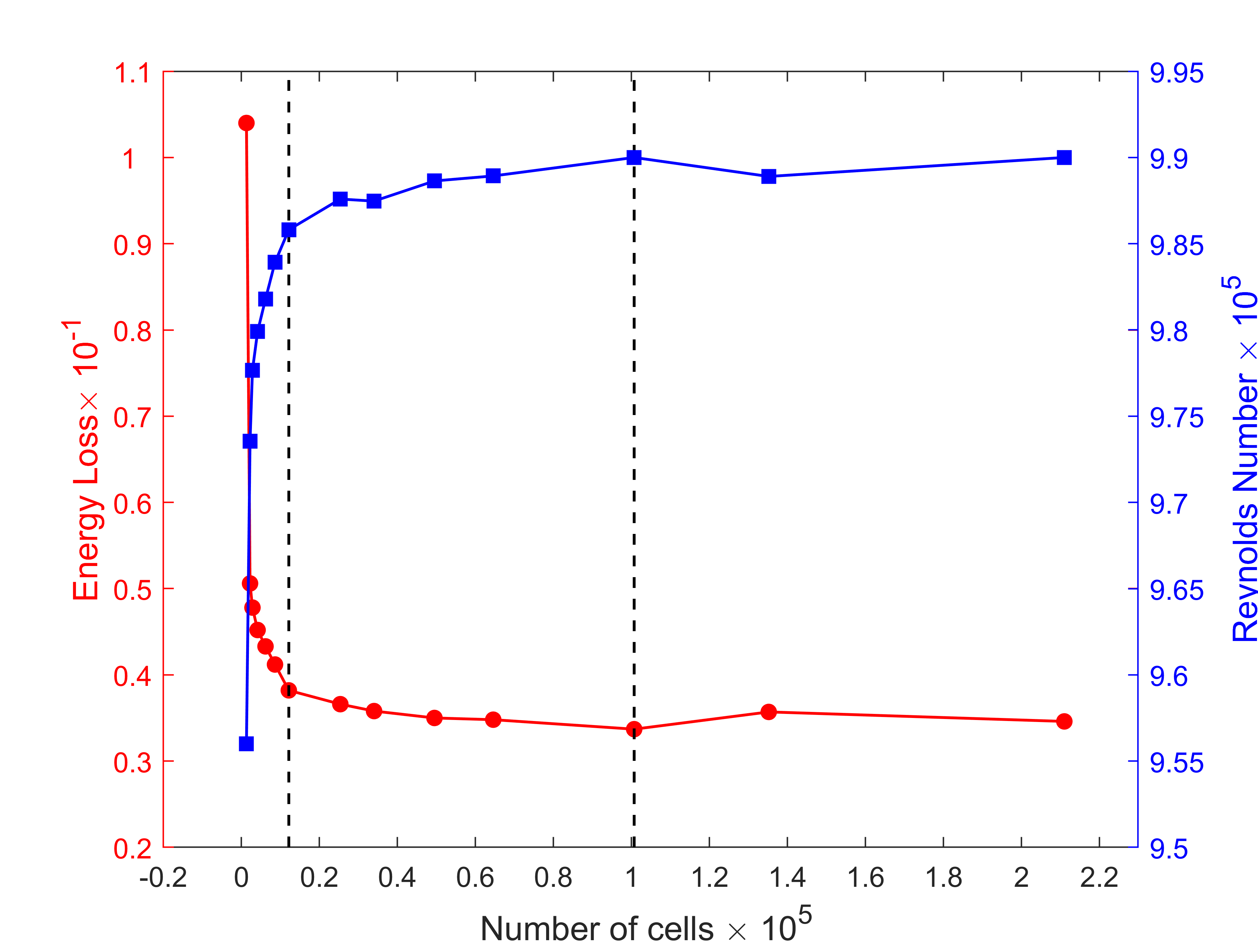}
\end{center}
\caption{Tests on grid independence for the engineering problem}
\label{fig:1-2}      
\end{figure}

\begin{table}[h]
\caption{Design conditions}
\begin{center}
\label{table_ASME}
\vspace{0.25cm}
\renewcommand\arraystretch{1.2}
\begin{tabular}{cl}
\hline
Condition name &      Value \\
\hline
inlet total temperature[$^\circ C$] &        709 \\
inlet total pressure[Pa] &     344740 \\
inlet flow angle[$^\circ$] &         90 \\
\text{Mach number at cascade outlet }[-] &    0.878 \\
\hline 
\end{tabular}  
\end{center}
\end{table}

\subsubsection{Results and Discussions}
For the turbine design optimization with seven design variables, we initially generate 42 LF samples and 21 HF samples by using the nearest neighbor sampling~\cite{guo_analysis_2018}. 
Different from the benchmark functions, the real optimal solution of is unknown, therefore we compare the performance of the different algorithms by fixing the total number of HF samples, i.e., the comparing algorithms will be terminated after consuming 35 HF samples in the iteration process. 
Considering the expensive cost of simulations, we run the engineering case five times to compare the performance of different algorithms. 

Before carrying out the optimization process, we use the leave-one-out validation (Loo-CV) to validate the accuracy of co-kriging, as shown in Fig.5. Table 7 shows the quantitative measures for the five tests. 
The definitions of cross-validated root mean squared error (CV-RMSE) and cross validated $R^2$ (CV-$R^2$) are shown in Eq.(16).
\begin{equation}
\begin{small}
\begin{split}
&\text{CV-RMSE}=\sqrt{\sum_{i=1}^{N}\left({y_{i}-\hat{y}_{i}}\right)^{2}/N}\\
&\text{CV-R}^{2}=1-\sum_{i=1}^{N}\left(y_{i}-\hat{y}_{i}\right)^{2} / \sum_{i=1}^{N}\left(y_{i}-\bar{y}\right)^{2}\\
\end{split}
\end{small}
\end{equation}

In Fig.5, the surrogate prediction values $\hat{y}$ and the true function $y$ are distributed closely to the line $y=\hat{y}$, which indicates that the co-kriging models we built were in good accuracy.
In Table 7, all the cross-validated $R^2$ are larger than 0.85, and most of them are larger than 0.95, which further confirms the good accuracy of co-kriging models. 
\begin{figure}[ht]
\begin{center}
\includegraphics[scale=0.75,trim=0 0 0 0]{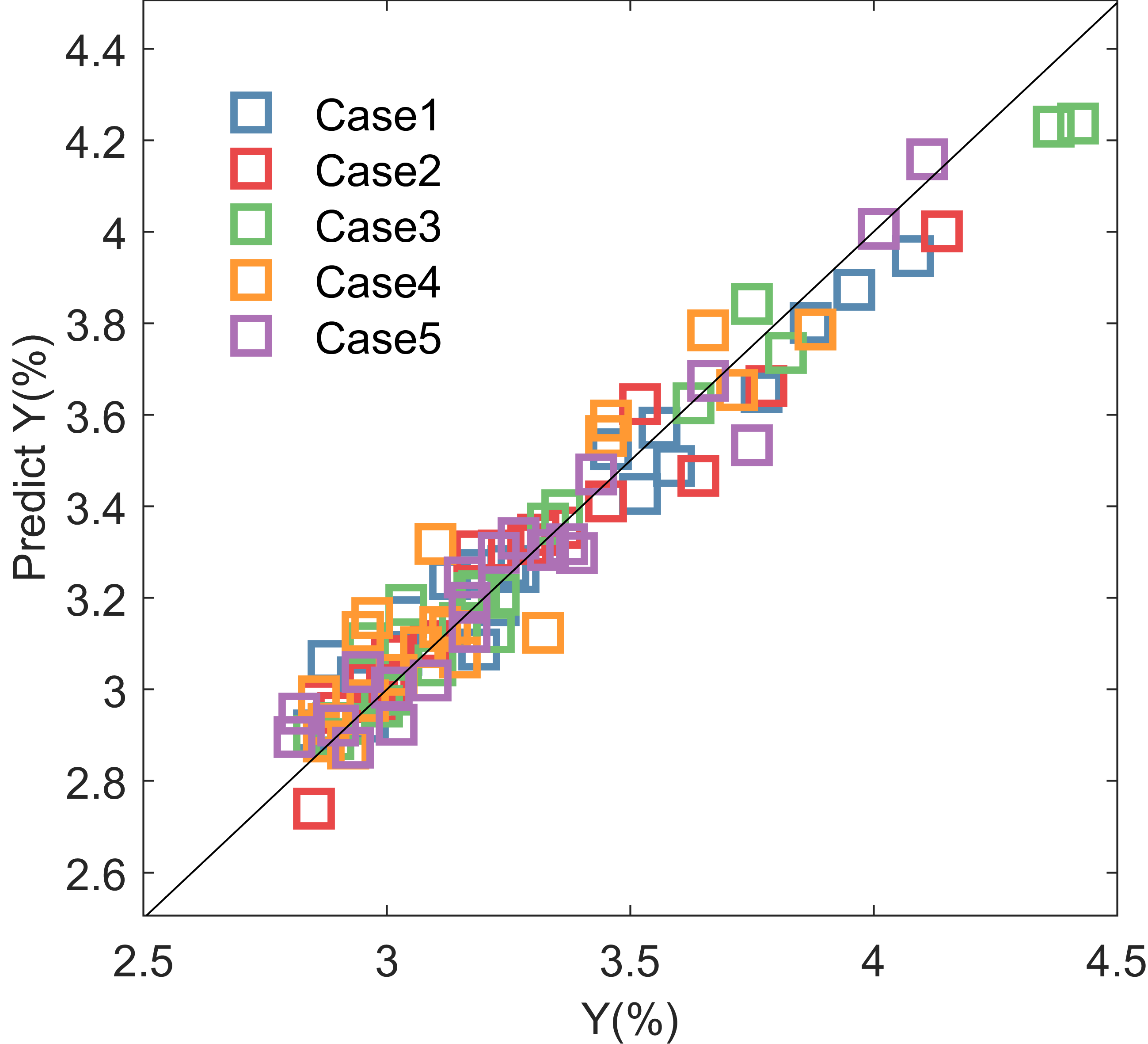}
\end{center}
\caption{Scatterplot of the cross validation results for the engineering problem}
\label{fig:1-3}      
\end{figure}

\begin{table*}[ht]
\centering
\caption{Cross-validated results of co-kriging model in the five tests for the engineering problem}
\begin{tabular}{ccccccc}
\hline
    Surrogate       &            &      Case1 &      Case2 &      Case3 &      Case4 &      Case5 \\
\hline
\multicolumn{ 1}{c}{co-kriging} &       CV-RMSE &   9.21E-02 &   1.12E-01 &   7.87E-02 &   3.03E-01 &   7.87E-02 \\
\multicolumn{ 1}{c}{} &        CV-$\text{R}^{2}$ &     0.938  &     0.952  &     0.970  &     0.862  &     0.953  \\
\hline
\end{tabular}  
\end{table*}

Table 8 shows the optimization results after consuming 35 HF samples, and Fig.6 shows the averaged convergence history of different algorithms. 
Similar to the results of benchmark functions, Filter-GEI and GEI-based MFO use significant less iterations to obtain the optimal solution.
More importantly, as shown in Fig.6, with the filter of HF samples, the Filter-GEI shows significantly faster convergence rate than the GEI-based MFO. The Filter-GEI also achieves the best optimal solution after consuming 35 HF samples. 
With the above, the effectiveness of Filter-GEI is further demonstrated, which can make full use of multi-fidelity and parallel computing resources to achieve the optimal solution most efficiently. 
\begin{table*}[h]
  \centering
  \caption{Optimization results after adding 35 HF samples}
  \renewcommand\arraystretch{1.2}
\begin{tabular}{cccccccc}
\hline
      \tabincell{c}{} &       Iteration&       Optimal(\%) &            Min(\%) &        Max(\%) &      \tabincell{c}{HF\\number}  & \tabincell{c}{LF\\number}\\
\hline
Filter-GEI &         $13\pm1$ &    $2.720\pm0.0245$ &      2.70 &     2.76 &      56 &$91\pm10$ \\
       GEI &          $8\pm0$ &      $2.760\pm0.0224$ &     2.73 &     2.79 &     56& $79\pm2$ \\
        EI &         $36\pm0$  &    $2.732\pm0.0192$ &      2.70 &     2.75 &      56& $77\pm0$ \\
     VF-EI &         $45\pm3$ &    $2.748\pm0.0239$ &     2.72 &     2.78 &      56& $51\pm3$ \\
augmented-EI &        $51\pm6$ &    $2.726\pm0.0358$ &     2.69 &     2.78 &      56& $59\pm9$ \\
\hline
\end{tabular}  
\end{table*}

\begin{figure}[ht]
\begin{center}
\includegraphics[scale=0.7,trim=12 0 0 0]{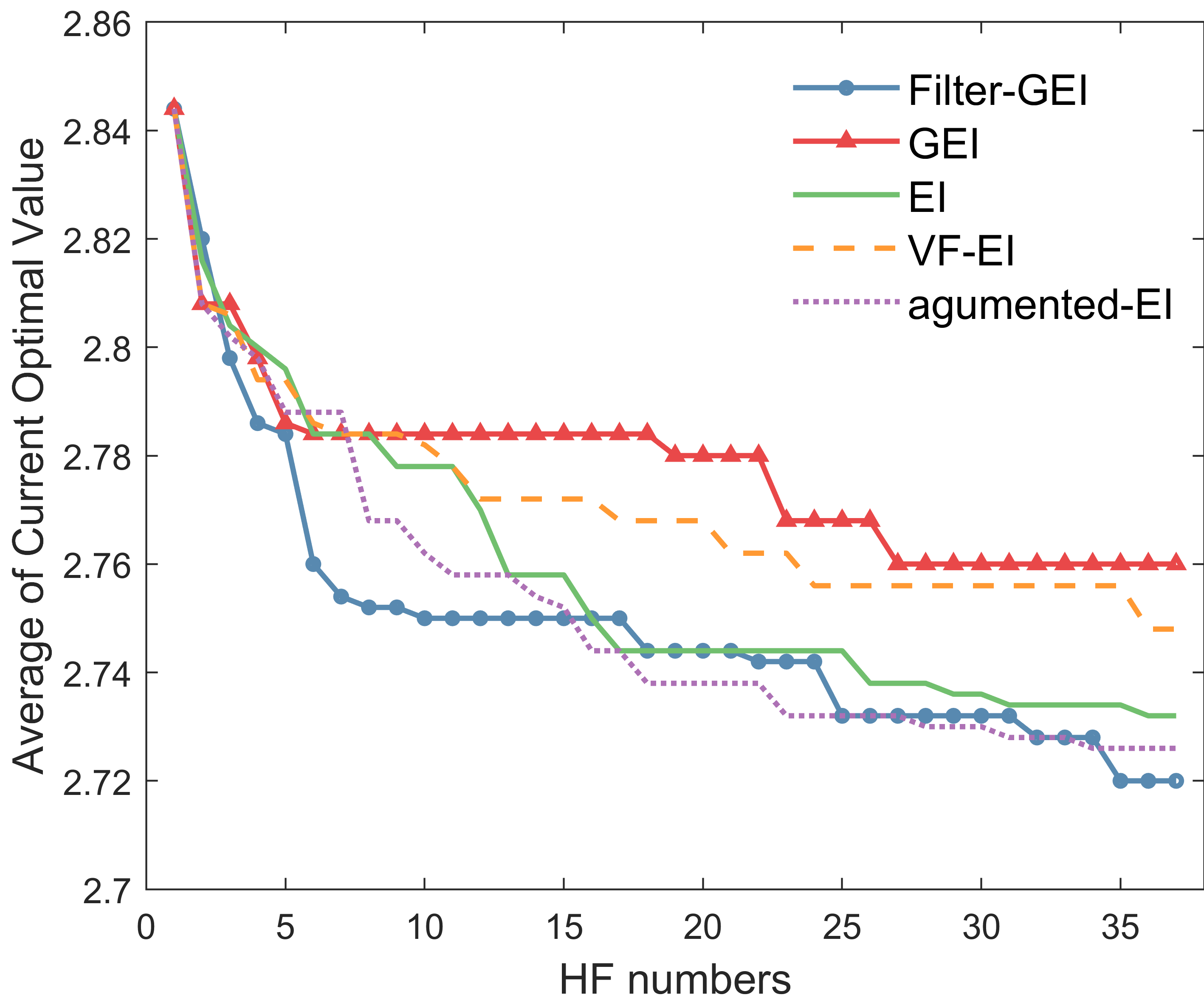}
\end{center}
\caption{Convergence history from the view of HF numbers}
\label{fig:1-4}      
\end{figure}
\section{Conclusion}
In this paper, we propose a new infill criterion named as Filter-GEI for multi-fidelity surrogate based optimization (MFO). 
Different from previous works, we propose to infill multiple high-fidelity (HF) and low-fidelity (LF) samples rather than one sample per optimization cycle, in order to make use of the multi-fidelity and parallel computing resources.
More specifically, considering the correlations between HF and LF simulations, we propose an adaptively filter function on top of the generalized expected improvement (GEI) acquisition function, and we name it as Filter-GEI.  
The Filter-GEI can adaptively filter the HF samples to be evaluated. Thereby, we can call more HF samples to exploit ”promising” areas with relatively better objective function value, while we query LF samples to explore the rest of areas globally. This way we achieve a good balance in between the local and global search by taking full advantage of multi-fidelity and parallel computing resources.
Through tests on five benchmark functions and one engineering problem of the turbine blade design, the proposed Filter-GEI outperforms the compared algorithms. 
Hence, the effectiveness of our proposed algorithm is well demonstrated.

\vspace{1cm}
\small{\noindent\textbf{Acknowledgements}\quad
The authors also would like to thank the anonymous referees for their valuable comments.}\\

\small{\noindent\textbf{Funding information}\quad
This work was supported by the National Natural Science Foundation of China (Grant No. 51676149).}

\section*{Compliance with ethical standards}
\small{\noindent\textbf{Conflict of interest}\quad
The authors declare that they have no conflict of interest.}\\

\small{\noindent\textbf{Replication of results}\quad
The code will be published once the paper is accepted.}

\nocite{*}
\bibliographystyle{unsrt}
\bibliography{FGEI}
\end{document}